\documentclass[pmlr]{jmlr}% new name PMLR (Proceedings of Machine Learning Research)

\usepackage{longtable}% for long tables
\usepackage{tikz-cd}
\usepackage{tikz}
\usetikzlibrary{positioning,arrows,automata,calc,shapes,shapes.geometric}
\tikzset{elliptic state/.style={draw,ellipse}}
\tikzset{initial text={}}

\usepackage{booktabs}
\usepackage[load-configurations=version-1]{siunitx} % newer version
\theorembodyfont{\upshape}
\theoremheaderfont{\scshape}
\theorempostheader{:}
\theoremsep{\newline}

\jmlrvolume{1}
\jmlryear{2022}
\jmlrworkshop{LearnAut 2022}

\title[Analyzing Büchi Automata with Graph Neural Networks]{Analyzing Büchi Automata with Graph Neural Networks}

  \author{
  \Name{Christophe Stammet} \Email{christophe.stammet@unifr.ch}\\
  \Name{Prisca Dotti} \Email{prisca.dotti@unifr.ch}\\
  \Name{Ulrich Ultes-Nitsche} \Email{uun@unifr.ch}\\
  \addr University of Fribourg, Switzerland
  \AND
  \Name{Andreas Fischer} \Email{andreas.fischer@hefr.ch}\\
  \addr University of Applied Sciences and Arts Western Switzerland, Switzerland
 }

\begin{document}

\maketitle

\begin{abstract}
Büchi Automata on infinite words present many interesting problems and are used frequently in program verification and model checking. A lot of these problems on Büchi automata are computationally hard, raising the question if a learning-based data-driven analysis might be more efficient than using traditional algorithms. Since Büchi automata can be represented by graphs, graph neural networks are a natural choice for such a learning-based analysis. In this paper, we demonstrate how graph neural networks can be used to reliably predict basic properties of Büchi automata when trained on automatically generated random automata datasets.
\end{abstract}
\begin{keywords}
machine learning, automata theory, infinite structures, dataset creation, büchi automata
\end{keywords}

\section{Introduction}
Büchi automata have been introduced in 1962 by J.R. Büchi \citep{buchi} as automata on infinite words and were shown to be a formalism accepting all the $\omega$-regular languages. In practice, they are used as models of reactive concurrent systems since they represent the indefinite running time of these systems very nicely.

Studying Büchi-automata leads to many different interesting theoretical questions, like minimization \citep{minim_ba}, complementation \citep{ba_compl} as well as a variety of practical applications, such as in security using greybox fuzzying \citep{fuzzy} or model checking \citep{mc, mc_ltl} by verifying a property, expressed as LTL (Linear Temporal Logic) formulas, on a Büchi automaton model of the system to verify.

With the development of machine learning, neural networks are playing an important part in many different tasks like speech recognition, image classification and many more. The importance of graph structures in many applications like biochemistry \citep{biochem}, social media analysis \citep{socmed} or applications in formal methods (like SAT solving \citep{maxsat}) lead to graph neural networks (GNNs) being a new tool to handle these many tasks which go beyond the scope of Euclidean space data. 

There are many different GNN architectures \citep{gnnsurvey}, which are constantly improving on the various problems posed by graphs. A few ventures into analysing automata on finite words have also been proposed (e.g. \citep{nn_dfa}), but Büchi automata on infinite words have to our knowledge not been covered in the GNN literature yet.

In this paper, we make a first step towards learning-based analysis of Büchi automata using GNNs. We focus on basic properties, such as emptiness of the accepted language, and generate random automata datasets for training, which are well-balanced with respect to the properties. In an experimental evaluation, we train standard GNN architectures on the synthetic data and demonstrate a promising performance on independent test sets.

The remainder of the paper is structured as follows. Section \ref{s2} introduces the basic definitions of Büchi automata and graph neural networks, Section \ref{s3} presents the random dataset generation for several basic properties of Büchi automata, and Section \ref{s4} details the experimental evaluation. Finally, we draw some conclusions in Section \ref{s5}.
\section{Preliminaries} \label{s2}
\subsection{Büchi automata}
There are different variants of automata on infinite objects. In this work, we will focus on non-deterministic Büchi automata on words (NBW) with multiple acceptance states. A NBW $\mathcal{A}$ is defined as a tuple $\mathcal{A} = (Q, \Sigma, \delta, q_{in}, F)$ where $Q$ is the set of states, $\Sigma$ is the alphabet, $\delta: Q \times \Sigma \longrightarrow 2^Q$ the transition function, $q_{in} \in Q$ the initial state and $F \subseteq Q$ the set of accepting states. A run $r$ on an infinite word $w = w_0 w_1 w_2 ...$ (where $w_i \in \Sigma$, $i \in \mathbb{N}$) of an automaton is defined as an infinite sequence of states $r = q_0 q_1 q_2 q_3 ...$ such that $q_i \in Q$ for $i \in \mathbb{N}$, $q_0 = q_{in}$ and $q_{i+1} \in \delta(q_{i}, w_i)$. Finite runs on finite words are defined similarly.

A state is called reachable if it is possible to find a finite run leading to that state starting at $q_0$. A state is called self-reachable if a non-empty finite run from that state to itself can be found. We define $\omega(r)$ as the set of states occurring infinitely often in an infinite run $r$.

A run $r$ on $w$ is called accepting if and only if $\omega(r) \cap F \neq \emptyset$. We then call $w$ accepted by $\mathcal{A}$. All the words accepted by $\mathcal{A}$ form the $\omega$-regular language $\mathcal{L}_\omega(\mathcal{A})$. Let $U_i$, $V_i$  (for $0 \leq i \leq n$) be regular languages, then we can write \citep{uvproof}
\begin{equation}
\label{uv}
\mathcal{L}_\omega(\mathcal{A}) = \bigcup_{i=0}^n U_i  V_i^{\omega}.
\end{equation}
This means that every word $w \in \mathcal{L}_\omega(\mathcal{A})$ is of the form $uv_0v_lv_2...$, where $u \in U_i, v_j \in V_i$ (for $0 \leq i \leq n$ and $0 \leq j$). Here, $u$ is a finite prefix and the $v_j$ are the looped paths of a self-reachable accepting state. 

\subsection{Graph Neural Networks}
A graph neural network (GNN) is a neural network which takes a graph structure as input. Using message passing, where node features are propagated to their neighbours, the GNN can perform either node classification, edge prediction or graph classification \citep{gnnsurvey}. Our datasets will focus on binary graph classification tasks, e.g. does the given NBW have a certain property or not.

The experiments in this paper will run using the most commonly used architecture on graph problems, notably Graph Convolutional Network \citep{gcn} layers, with each layer passing the node features to its neighbours as follows:
\begin{equation}
H^l = \sigma ( D^{-1/2} A D^{-1/2} H^{l-1} W^{l-1} ),
\label{gcnlayer}
\end{equation}
where $H^l$ is the hidden state of layer $l$ (i.e. the initialized node features if $l=0$), $A$ the adjacency matrix, $D$ the degree matrix, $W^l$ the trainable weight matrix of layer $l$ and $\sigma$ the activation function (we use the rectifier linear unit (ReLU) function: $ReLU(x) = max\{0,x\}$).

The GNN for classifying automata is trained on a labelled dataset with each data element representing one NBW with a binary label denoting whether it satisfies the given property or not. The creation of these datasets will be described in the next section.

\section{From Büchi automata to datasets} \label{s3}
This section will highlight how the NBW are created and how they are encoded to fill the datasets for neural network training. We use the PyTorch Geometric library for deep learning on graphs \citep{pytorch_geom} as framework for training the neural networks, thus using the representation of our automata as \textbf{Data} instances as defined by the \textbf{Dataset} classes in PyTorch Geometric.
\subsection{Random generation and encoding of Büchi Automata}
The random automata generation is based on the Erdős-Rényi graph model \citep{erdosrenyi}, where a graph $G(n, p)$ is defined as a graph with $n$ nodes and all possible edges are included with probability $p$. To extend this approach to NBW, it suffices to count all possible edges of the graph structure once for each symbol in $\Sigma$. In addition, a second probability $p_{acc}$ is defined to determine for each state if they belong to $F$, i.e. are accepting.

For the NBW to be able to be handled by the GNNs, both the transition labels and the node features have to be encoded as vectors, which leads to one of the novel ideas this paper presents in encoding the automata for a neural network dataset. The symbols of $\Sigma$ will be encoded as one-hot vectors (of length $\vert \Sigma \vert$) and will be the labels for the transitions. The feature vector of each node will encode the type of the node (initial state or accepting states) as binary flags in each nodes feature vector (where the first element will encode if the node is the initial node and the second if it is accepting or not). 

Furthermore, we will introduce a number of additional elements in each node feature vector that will be used by the GNN to store additional information of the automaton's structure it learns during training. These additional label elements can be initialized either to zero, to $0.5$ or a random value in $[0,1)$. The amount of these leads to a new parameter for the automaton encoding $n_{add}$ denoting the number of these additional elements.

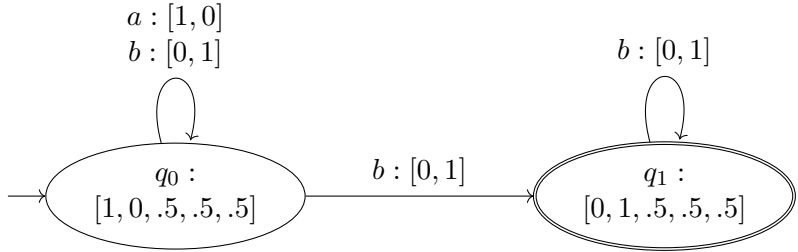
\begin{figure}
\centering
\begin{tikzpicture}[node distance=2cm,on grid,auto]
\node [elliptic state, initial, align=center] (q_0) {$q_0:$\\$[1,0,.5,.5,.5]$};
\node [elliptic state, accepting, align=center] (q_1) [right=6.5cm of q_0] {$q_1:$\\$[0,1,.5,.5,.5]$};

\path[->]
(q_0) edge [loop above] node[align=center] {$a:[1,0]$\\$b:[0,1]$} (q_0)
      edge node {$b:[0,1]$} (q_1)
(q_1) edge [loop above] node {$b:[0,1]$} (q_1);
\end{tikzpicture}
\caption{Example of a vector encoding on automaton}
\label{a2}
\end{figure}

An example automaton (accepting all $\omega$-words over the alphabet $\Sigma = \{a,b\}$ containing finitely many $a$'s) and its respective encoding of the node features and transition labels can be seen in Fig. \ref{a2}, where $n_{add} = 3$ with these features being initialized to 0.5. 

\subsection{Dataset creation}
The randomly generated automata yield the graph information for each element of the dataset. For training the GNN, the data elements require a label denoting whether they possess the specific automaton property or not. To create a few different datasets on simple tasks on NBW, the following properties on an automaton (all over the alphabet $\Sigma = \{a,b\}$) are used to create different datasets:
\begin{itemize}
\item \textbf{is\_empty:} This verifies if the $\omega$-language the automaton accepts is empty or not. Structurally, for an automaton to be empty, it has to either contain no accepting states at all, have no accepting states reachable from the initial state or requires all the reachable accepting states to not be self-reachable.
\item \textbf{min1\_b:} This verifies if the given automaton accepts an $\omega$-word containing at least one $b$. 
\item \textbf{inf\_b:} Here, the given automaton is classified by whether it accepts $\omega$-words containing infinitely many $b$'s or not. In order to accept such a word, one edge in the loop of a self-reaching accepting state has to read a $b$.
\end{itemize}
Checking for each of the 3 properties is algorithmically simple, but differs in their requirement of analysis of the automaton structure. To illustrate this, let us look at the structure of each $\omega$-word accepted by an automaton being of the form $w = uv_0v_1v_2...$ (for $u \in U_i, v_j \in V_i$ for $0 \leq i \leq n$, $0 \leq j$ with $U_i$, $V_i$ as seen in Equation \ref{uv}). 
\begin{itemize}
\item \textbf{is\_empty} is a graph reachability problem and doesn't require the transition information to be analyzed (i.e. an existence of any accepting $\omega$-word $w$ satisfies the property)
\item \textbf{min1\_b} requires that at least one transition leading to an accepting state (that is also self-reachable) or one of the self-reachable loops reads a $b$, i.e. an accepting $w = uv_0v_1v_2...$ requires a $b$ in either $u$ or at least one of the $v_j$ ($0 \leq j$)
\item \textbf{inf\_b} requires at least one $b$ transition to be in infinitely many loops of a self-reachable accepting state, i.e. satisfying the property requires a $b$ in infinitely many $v_j$ ($0 \leq j$), for an accepting $w = uv_0v_1v_2...$.
\end{itemize}

Another big part of the dataset creation process is balancing the dataset to contain all possible cases equally often. For the classifying label, this simply means that half the data from the dataset should satisfy the respective property and half should not.

However, inside these classes, the structures of the NBW can vary greatly and due to the random generation, not all of these different structures may occur equally often. Without addressing this, the dataset would train mainly on the more regularly occurring structures and dismiss those that are less likely to be generated randomly.

To illustrate this, let us start by looking at the property of emptiness, i.e. if a given automaton accepts at least one $\omega$-word or not. For a NBW to accept a non-empty $\omega$-language, at least one of the accepting states has to be both reachable and self-reachable. Thus, for a NBW to not accept any words, the automaton has to either
\begin{itemize}
\item not have a single accepting state
\item none of the accepting states are reachable
\item none of the reachable accepting states are self-reachable.
\end{itemize}
To guarantee a balanced dataset, our dataset creation ensures that all these sub-classes appear equally often.

In addition to that separation into sub-classes for the emptiness problem, we are also ensuring that, for all the properties treated in this paper, the length of the self-reachable path of an accepting state of an automaton satisfying the given property is balanced as well. Since the random creation would strongly balance the dataset towards the case where the length of such an accepting cycle is 1 (i.e. a self-loop over an accepting state), we guarantee in the datasets that the automata satisfying the property contain an equal number of data elements where this self-reachable path length is 1, 2 and 3 or more.

\section{Experimental Setup and Discussion} \label{s4}
This section will show the results of the classification accuracy over the different property datasets using a simple GCN architecture. We will start by presenting the different parameters and then show how the neural network learns over the various datasets. These classification accuracies will then be discussed in the second part of this section.

For reproduction of paper results or for trying out different dataset purposes, the codes for dataset creation and neural network training can be found online\footnote{\url{https://github.com/StammetC/BuchiAutomata_for_GNN}}.
\subsection{Experiment setup}
As described in Section 3.2, we are looking at datasets encoding 3 different problems on NBW. To create the datasets for each property, there are many different parameters to adjust:
\begin{itemize}
\item \textbf{automaton size $n^{min}, n^{max}$:} The range of size (number of nodes) of each automaton.
\item \textbf{edge probability $p^{min}, p^{max}$:} The probability range for any possible edge to exist in the automaton.
\item \textbf{acceptance probability $p_{acc}^{min},p_{acc}^{max}$:} The probability range of a node to be accepting.
\item \textbf{alphabet size $s = \vert \Sigma \vert$:} The number of symbols in the alphabet. 
\item \textbf{empty feature length $n_{add}$:} The amount of additional elements in each node feature vector.
\item \textbf{empty feature initialization $0$, $0.5$ or $random$:} The value to which the additional node feature vector elements are initialized.
\item \textbf{dataset size $d$:} The number of automata in the dataset.
\end{itemize}
For readability and simplicity, we tried to fix most parameters for this section to highlight the results only for some important parameter changes, i.e. for all results in this section, we have set $p^{min} = 0.1$, $p^{max} = 0.3$, $p_{acc}^{min} = 0.1$, $p_{acc}^{max} = 0.15$, $s = 2$ (i.e. for all our NBW: $\Sigma = \{a,b\}$), 
\begin{figure}
\centerline{\includegraphics[width=12cm]{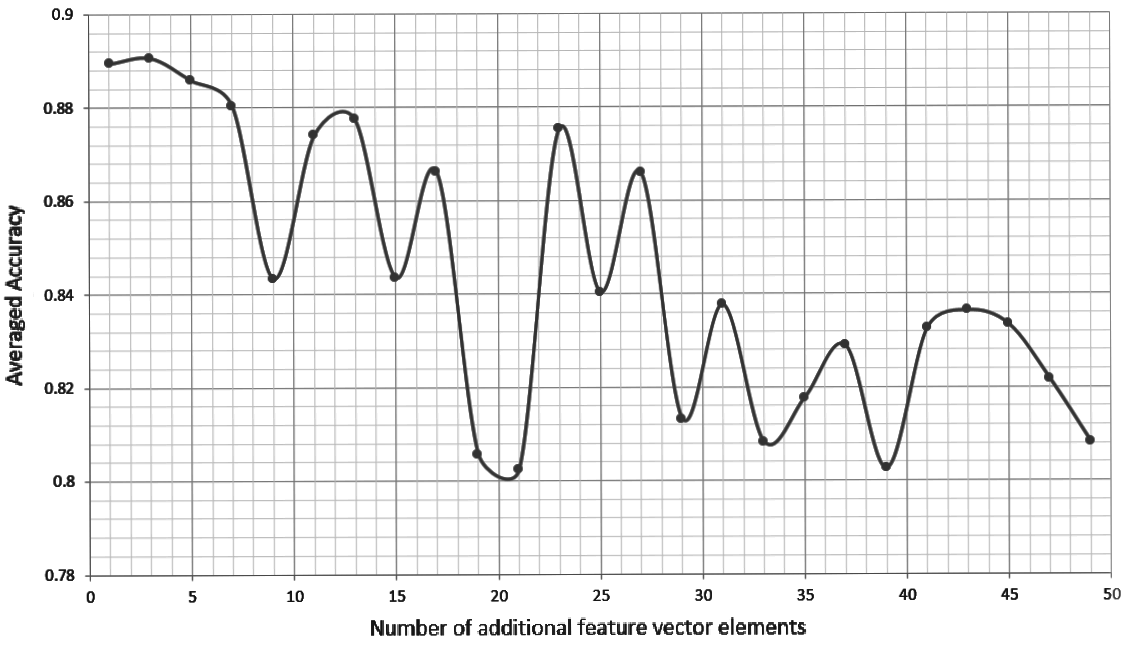}}
\caption{Average accuracies for different $n_{add}$ values}
\label{nadd}
\end{figure}
$n_{add} = 3$ (c.f. ''Fig. \ref{nadd}'' for averaged accuracies over different dataset properties) and empty features are initialized to $0.5$.\\
We refer to the balanced datasets as $property\_d\_n^{min}\_n^{max}$, where $property$ is either $empty$, $min1b$ or $infb$, e.g. $min1b\_1000\_3\_9$ is the dataset containing 1000 NBW with $3 \leq n \leq 9$ nodes classifying each automaton whether it accepts an $\omega$-word containing at least one $b$ or not.

The neural network used for all experiments consists of 3 GCN Conv \citep{gcn} layers with the ReLU activation function, each layer with the added self-loop option disabled and 20 hidden channels\footnote{Different amounts of hidden channels were tested here, with 20 yielding the highest accuracies.}. After the layers, for classification, we use a mean pooling layer and a simple linear classifier for the final classification. The optimizer is \textbf{adam} \citep{adam}, with a learning rate of $0.01$ and the used loss function is \textbf{categorical crossentropy}. Every GNN is trained for 75 epochs and the data is aggregated together in batches of size 125. 
\subsection{Results and discussion}
\begin{table}[]
\centering
\caption{Classification accuracies (in $\%$) using various datasets}
\begin{tabular}{c|cc}
                                       & \multicolumn{2}{c}{\textbf{Test set}$^{\mathrm{a}}$}                       \\
\textbf{Training set}                  & \textit{\textbf{500\_3\_9}} & \textit{\textbf{500\_10\_25}} \\ \cline{2-3} 
\textit{\textbf{infb\_250\_3\_9}}    & 89.7 $\pm$ 0.6              & 89.2 $\pm$ 0.7                 \\
\textit{\textbf{infb\_1k\_3\_9}}     & 89.4 $\pm$ 0.7              & 93.1 $\pm$ 0.7                \\
\textit{\textbf{infb\_10k\_3\_9}}    & 90.2 $\pm$ 0.4              & 93.9 $\pm$ 0.9                \\
\textit{\textbf{infb\_50k\_3\_9}}    & 91.0 $\pm$ 0.5              & 95.1 $\pm$ 0.8                \\
\textit{\textbf{min1b\_250\_3\_9}}   & 87.1 $\pm$ 0.9              & 87.6 $\pm$ 0.8                \\
\textit{\textbf{min1b\_1k\_3\_9}}    & 88.8 $\pm$ 0.8              & 87.1 $\pm$ 1.1                \\
\textit{\textbf{min1b\_10k\_3\_9}}   & 89.7 $\pm$ 0.7              & 86.7  $\pm$ 1.6               \\
\textit{\textbf{min1b\_50k\_3\_9}}   & 90.9 $\pm$ 0.4              & 91.6 $\pm$ 2.0                \\
\textit{\textbf{emptiness\_250\_3\_9}} & 79.3 $\pm$ 4.1              & 63.4 $\pm$ 7.9                \\
\textit{\textbf{emptiness\_1k\_3\_9}}  & 87.2 $\pm$ 2.2              & 85.4 $\pm$ 1.4                \\
\textit{\textbf{emptiness\_10k\_3\_9}} & 88.9 $\pm$ 3.7              & 87.1 $\pm$ 3.2                \\
\textit{\textbf{emptiness\_50k\_3\_9}} & 90.4 $\pm$ 2.8              & 90.9 $\pm$ 2.8    \\
\hline
\multicolumn{3}{l}{$^{\mathrm{a}}$ Each test set property is the same as its respective training set.}
\end{tabular}
\label{accs}
\end{table}
The accuracies over the various testing datasets given their respective training datasets (in the first column) are shown in ''Table \ref{accs}''. Given is the mean accuracy and standard deviation over 10 training runs. 

The first thing that can be spotted is that, for all properties, the accuracies are generally improving if the training dataset is larger, i.e. more different data to learn on is available, which reflects how the learning is expected to behave. Something interesting to note is that, for the largest datasets of all properties, the classification performs better on the test sets containing slightly larger automata than the training set. This can be explained by the simple nature of the structures that need to be recognized for classification (e.g. the looped paths of self-reaching accepting states), which are more efficiently found in the larger, sparser (due to the nature of the random generation) automata.

Especially for the infinitely many $b$ property, we can see that the neural network is very good at generalizing during the learning, with the accuracies being consistently higher on the testing datasets containing larger automata than in the training set. This generalization, where the neural network is tested on larger automata than it is trained on, illustrates very nicely that these neural networks are able to understand the structural specificities of these properties when training on sufficiently large sets, thus encountering most of the structural possibilities during training.

Another sign of this is looking at the emptiness property, where on the small training dataset, the neural network fails to generalise for the larger automata. This can be explained that during training on the small dataset, the neural network does not encounter enough of the possible structures to generalise, which is more important for the emptiness property than for e.g. infinitely many $b$ property, which can be classified by focussing more locally on the accepting states.

Overall, the classification accuracies that were reached during these experiments showed us that the neural network architecture that was used was able to learn about the automata structures and, especially with the architecture being very simple and unoptimized, these are good results to start improving in different directions, which we will cover in the next section.

\section{Conclusion and Future Work} \label{s5}
In this paper, we have proposed a learning-based approach to analyze Büchi automata using GNNs. It was demonstrated that by means of generating random automata, GNNs are able to learn basic properties, such that they can be predicted on independent test automata with high precision. In particular, it is noteworthy that the prediction was also successful when the test automata were significantly larger than the automata used for training, demonstrating a strong generalization capability of the proposed GNN-based approach.

There are several promising lines of research for future work. First with respect to the GNN architecture. This paper focusses on creating the datasets and showing their potential. Due to this, having focused on standard architectures in our experiments, we would expect that exploring more network architectures (e.g. one better suited for edge feature propagation) and going deeper in terms of message passing steps can improve the prediction performance. Analysis of the node features, especially the added elements, could also yield insight into the neural networks learning process.

Secondly, regarding the tasks on NBW. As a next step, we aim to address more challenging problems related to Büchi automata. Concretely, a goal would be to have neural networks tackle the universality problem of Büchi automata, which is a property of a NBW which is satisfied if and only if the automaton is accepting every possible $\omega$-word over its alphabet. This property check is computationally difficult ($2^{\mathcal{O}(n\text{ log }n)}$ as shown by Safra in \citep{safra}) and the complexity bottleneck for the complementation of Büchi automata, a vital problem in program verification.
 
Finally, an important open question relates to the transferability of learned GNN weights from one type of problem to another. Ideally, performing pretraining tasks on basic properties of Büchi automata can be used as an initial step to initialize the network weights before fine-tuning the network to more challenging tasks. Specifically the aforementioned universality problem for Büchi-automata will be an interesting test case with respect to pretraining: As universality of a Büchi automaton is equivalent to the emptiness of its complement, training a GNN to recognise emptiness, which we have covered in this paper, appears to be particularly well suited as a pretraining prior to training for detecting universality.

\bibliography{gnnpaperbib}

\end{document}